# Artificial intelligence approaches for energetic materials by design: state of the art, challenges, and future directions

Joseph B Choi[1], Phong C. H. Nguyen[1], Oishik Sen[2], H. S. Udaykumar[2,*], Stephen Baek[1,3,*]

[1]University of Virginia, School of Data Science, Charlottesville, VA 22903, United States
[2]University of Iowa, Department of Mechanical Engineering, Iowa City, IA 52242, United States
[3]University of Virginia, Department of Mechanical and Aerospace Engineering, Charlottesville, VA 22903, United States
*Corresponding Authors: hs-kumar@uiowa.edu, baek@virginia.edu

**Abstract:** Artificial intelligence (AI) is rapidly emerging as a enabling tool for solving complex materials design problems. This paper aims to review recent advances in AI-driven materials-by-design and their applications to energetic materials (EM). Trained with data from numerical simulations and/or physical experiments, AI models can assimilate trends and patterns within the design parameter space, identify optimal material designs (micro-morphologies, combinations of materials in composites, etc.), and point to designs with superior/targeted property and performance metrics. We review approaches focusing on such capabilities with respect to the three main stages of materials-by-design, namely representation learning of microstructure morphology (i.e., shape descriptors), structure-property-performance (S-P-P) linkage estimation, and optimization/design exploration. We leave out "process" as much work remains to be done to establish the connectivity between process and structure. We provide a perspective view of these methods in terms of their potential, practicality, and efficacy towards the realization of materials-by-design. Specifically, methods in the literature are evaluated in terms of their capacity to learn from a small/limited number of data, computational complexity, generalizability/scalability to other material species and operating conditions, interpretability of the model predictions, and the burden of supervision/data annotation. Finally, we suggest a few promising future research directions for EM materials-by-design, such as meta-learning, active learning, Bayesian learning, and semi-/weakly-supervised learning, to bridge the gap between machine learning research and EM research.

## 1. Introduction

Energetic materials (EM) cover a wide spectrum of propellants, pyrotechnics, and explosives and are key components in military applications for propulsion and munition systems and in civilian applications such as construction and mining [1]. Heterogenous/composite EMs have complex microstructures which significantly influence—along with chemistry—the property and performance of these materials [2-8]. There is increasing research interest in controlling the microstructure of EM, to engineer their properties and performance for targeted functional specificity [9-10].

EMs are typically solid-solid composites of organic energetic crystals (commonly CHNO compounds), inclusions (i.e., metals, nanoparticles), and plastic binders. The CHNO materials are commonly categorized based on how sensitive they are to an external load/mechanical insult. They can range from 'insensitive' (such as TATB-based EMs [11]) to 'highly sensitive' (PETN-based EMs [12-13]) with others such as HMX, CL-20, and RDX ranging in between [14]. The sensitivity is closely connected with the molecular structure of these species of EMs within the CHNO family. However, when they are formed into propellants and explosives, the sensitivity is also impacted by the physical structure, composition, and formulation of the material mixtures, as reviewed by Handley et al. [1]. In other words, the design of a mixture and its microstructure can define the overall properties and performance characteristics of formed EM, thus opening the possibility of systematic methods to engineer materials by their design.

## 1.1 Materials-by-design (MbB) for EM

Traditional practices for EM selection, formulation, and design, however, take a "cut-and-try" approach, in which experiments are limited to small regions of the design parameter space. Hence, the discovery of a new design is not only driven by strong human intuition and domain expertise, but also by serendipity. In addition, such Edisonian approaches are laborious, expensive, and time-consuming and thus it becomes practically infeasible to fully explore the design space and discover an optimal design. Rather, only a certain limited design space is likely to be explored due to various human factors and legacy knowledge which restrict the designer within familiar, previously explored design subspaces. Furthermore, the inherently hazardous nature of EM limits exploration and leads to long development lead times.

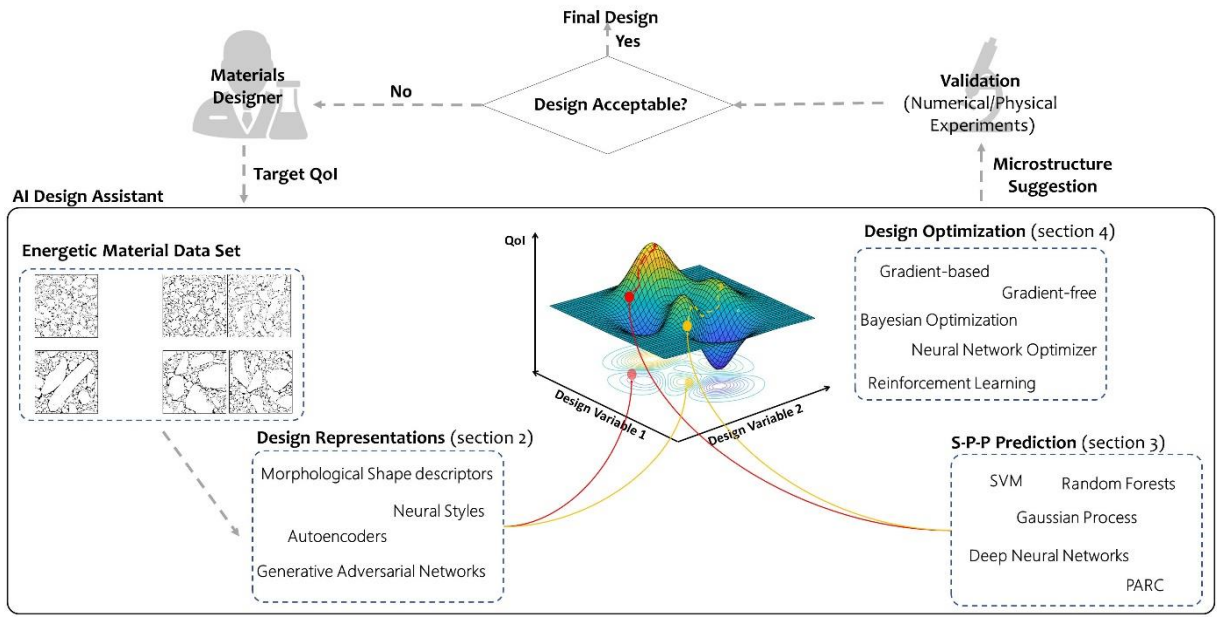

**Figure 1:** Overview of the materials-by-design (MbD) framework. The MbD framework is an iterative process where the materials designer specifies a target QoI, AI design assistants suggest a microstructure for the targeted QoI, each suggestion is validated by numerical/physical experiments, and repeats until the suggested design is acceptable to the materials designer. The MbD framework can be broken down into three subtasks (commonly observed throughout the field of material science), such as (1) design representation to define the design search space (Section 2), (2) establishing S-P-P linkage to estimate the property reliably and rapidly (Section 3), and (3) efficient optimization method as evaluating all the design candidates in the vast search space is impractical (section 4). The different AI methods are applied to each of the subtasks to assist the MbD framework to converge to an optimal design in a practical amount of time.

     The development of computational models has, to some extent, alleviated these problems by accelerating the evaluation of the property and performance of different EM microstructures. However, accurate computational models commonly take several hours or days (in the case of continuum/coarse-grain models [15-16]) to weeks (in the case of atomistic scale models [1]) to execute on high-performance computing clusters (HPC) [1, 17-19]. In addition, computational models are only as good as the physics sub-models (constitutive models and closure laws) that are built into them. Even for the most widely used EM species, there are gaps in knowledge (e.g., in material properties, material strength models, reaction chemistry models, etc.) [1] and establishing physically correct material descriptions is an area of active research [20]. In the case of EM, material properties and models can be difficult to obtain due to several reasons, such as anisotropy of the crystal, the sensitivity of the crystal to loading intensity and direction, strong dependence on pressure and temperature, the tendency of such materials to undergo phase transitions, and the hazardous nature of the materials.

     While the above challenges continue to exist in the design and formulation of EMs, there are new developments that offer routes to accelerated design of improved materials. In particular, recent advancements in manufacturing processes, such as additive manufacturing/3D printing [4, 21-22], enable better control and more degrees of freedom in synthesizing microstructures. Also, the improvement of computational models and algorithms allow for virtually prototyping microstructure designs to facilitate the exploration of the EM design space [23]. If the combination of new manufacturing approaches and in silico design can be orchestrated, the possibility of architected microstructures that can lead to precise control of sensitivity and performance can be realized. This paradigm of

"materials-by-design" (hereinafter abbreviated as MbD) relates to the task of designing material structures for a targeted set of properties (or quantity of interest, denoted by QoI), with iterative evaluations of material design candidates [24], as illustrated in Figure 1. Here, note that MbD is a subset of materials design, in which material structures are of particular interest to be optimized. Efficient MbD would require a rapid method for establishing structure-property-performance (S-P-P) linkages and an effective strategy to minimize the number of costly (numerical and physical) experiments on the way to an optimal target material in possibly a vast search space. In the past decade, the rapid growth of artificial intelligence (AI) and machine learning (ML) has opened the possibility of "learning" S-P-P linkages from simulation and experiment data, as well as "exploring" optimal designs through data-driven design exploration methods.

## 1.2 Taxonomy and terminology of AI/ML

Broadly speaking, AI can be viewed as a superset of ML encompassing general techniques to enable computers to emulate human thoughts and perform tasks on its own. ML, on the other hand, refers to methods for achieving such "intelligence" by assimilating trends and patterns from data. For example, in the context of MbD, ML could serve the role of assimilating mathematical relationships among process, microstructure, property, and performance parameters (Section 3). ML can also serve the purpose of discovering latent statistical patterns in trends in material data and producing statistically representative descriptors for different materials (Section 2). AI, on the other hand, could serve the role of amalgamating these different ML models, towards the goal of finding optimal design candidates via self-experimentation and/or human collaboration (Section 4). Note that, despite such a taxonomical distinction between AI and ML, we will make use of these terms interchangeably in this paper: 'AI/ML' as a collective term that refers to data-driven computational methods.

Another taxonomical label important for the discussion in this paper is deep learning (DL). DL is a subset of ML and refers to specific ML methods based on deep neural networks [25]. The word "deep" comes from the fact that there is more than one hidden layer of neurons stacked sequentially in a network, in order to make the model more capable of learning complex patterns and behavior. In particular, convolutional neural networks (CNN) are one of the popular deep neural network methods. Unlike other neural networks, CNNs employ "local connectionist" architecture [26], where the artificial neurons are connected to elements within a neighborhood (e.g., 3x3 area of pixels). Such local connectionist architectures reduce the computational burden by pruning global connections, enabling "going deeper." In addition, locally limited connections allow for the learning power of CNNs to be focused on smaller receptive fields, simplifying the learning task for individual neurons. Yet, CNNs can capture global features effectively through stacked CNN layers, in which earlier layers capture local geometric primitives such as edges, blobs, corners, etc., while deeper layers capture global geometric features ("bigger picture") by combining such local primitive features passed from the earlier layers.

## 1.3 Strengths and challenges of AI/ML approaches

CNNs, and more generally DL methods, have undoubtedly played a critical role in the recent advancement of computer and machine vision. With CNNs, computers can now classify images and videos, detect and recognize objects in them, delineate boundaries, generate text image captions, and so on [27-28]. They can also generate realistic-looking images [29], interpolate/extrapolate unseen views [30], enhance resolutions [31], reconstruct an original image by removing cracks and artifacts [32], and many other visual cognition tasks that were previously impossible. Such advances in the computer vision community have also created new opportunities in the applied physics and materials science community, which has acknowledged information and big data science as a "fourth paradigm" of material science [33]. While data-driven methods for material design and discovery are in their early stages of research, a rapidly growing body of AI/ML techniques has been applied in recent years covering a wide range of applications across length and time scales, from atomistic-scale molecular compounds [34-38] to macro-scale composite design [39].

However, there are also concerns around AI/ML in materials research, questioning the practical impact and efficacy of those methods. One significant concern could be the 'data hungriness' of AI/ML models. Typically, AI/ML methods, especially modern deep learning methods, expect a large quantity of labeled data covering diverse conditions. For example, the computer vision revolution centered around CNNs was only possible because of a large collection of human-annotated image data set called 'ImageNet' comprised of 1.4 million natural images with 1,000 different object class labels manually annotated by human annotators [40]. In contrast, material data are usually costly and laborious to acquire, requiring hours and days of numerical computations or lab experiments.

In addition, even if there was a sufficiently large number of datasets, computer vision algorithms right out of the box may not be practical for material science problems. Instead, as Nguyen et al. [41] argue, AI/ML models must be informed with governing equations, boundary conditions, and other physical constraints. This creates a critical need for materials-specific research and development on model architectures, requiring the examination of training objectives, training/testing protocols, and problem formulations that best utilize the power of AI/ML.

Furthermore, AI/ML methods in materials research must be explainable or "interpretable", so that new trends and patterns discovered by these methods can be interpreted and distilled into new knowledge that can eventually be added to the existing body of knowledge/practice. However, many AI/ML methods are not necessarily explainable, especially in the case of modern DL algorithms. Rather, they tend to work like a 'black box' in which there is no comprehensible reasoning or justification available as to why the model sees certain input-output relationships. For this reason, critics of AI/ML in materials research argue that these new methods are not trustworthy, question their scientific value to the research community and the generalizability beyond the materials and prediction tasks they are trained for.

## 1.4 Scope of the paper

With this background of promises and challenges, this review paper aims to provide a perspective view of the ways in which AI/ML methods have been used and adapted within the materials research community and their potential role and impact in the future. There are many exciting frontiers for AI/ML in materials science, but we will limit the scope of this review to AI/ML methods for the MbD of EMs. Specifically, our primary focus is on heterogeneous/composite EMs covering propellants, pyrotechnics, and explosives. In terms of the length-scale, we limit our scope to the physics of the geometrical features (i.e., distribution of crystals, grains, inclusions, defects, etc.) to the property and performance of the materials by focusing on meso-scale and macro-scale problems. Therefore, the application of AI/ML methods to connect atomic and molecular features is excluded from this review as there are other review papers specifically on the atomic/molecular scale aspects already [34-38]. On the other hand, data-driven methods for inert porous media (e.g., sandstone) are included as it has a similar pattern to the data with the energetic materials that are porous, complex, and heterogeneous. Finally, we limit the scope only to the S-P-P linkage and leave out the process in the Process-S-P-P (P-S-P-P) linkage as there are very few papers on this subject and much work remains to be done to establish this connectivity.

The remainder of the paper is organized as follows. In Section 1.5, we first describe the survey methodology adopted herein. We then define the problem of MbD in Section 1.6, by breaking it down into three subtasks, namely (1) microstructure representation to define a search space, (2) S-P-P linkage to solve the forward problem, and (3) optimization methods to efficiently navigate the design search space. The opportunities and challenges for inverse design for the materials science community in general and specific to the materials in scope are also discussed in Section 2 – Section 4. The challenges for AI/ML-driven EM research and possible directions of investigation are discussed in Section 5, followed by the conclusions in Section 6.

## 1.5 Survey methodology

We reviewed relevant papers in the field with publication dates ranging between 2000 and 2022. Articles published after 2012 were weighted more, to understand more recent trends. We mainly focused on papers from reputed journals and publishers such as Science, Nature, IEEE, Elsevier, Springer, and Wiley. When necessary, especially for the sake of probing the newest trends, some papers have also been selected from preprint repositories such as ArXiv. From these sources, we reviewed various applications of data-driven methods in energetic materials. The keywords used as search criteria were ("data-driven"), ("optimization"), ("material-by-design"), ("design"), ("approximation"), ("inverse design"), ("material discovery"), ("porous media"), ("data-driven" OR "deep learning" OR "computational design" OR "computational modeling OR "neural networks" OR "computer simulation") AND ("materials by design" OR "design") AND ("energetic materials" OR "pyrotechnics" OR "explosives" OR "propellants" OR "porous media").

As aforementioned, this review focuses on AI/ML-related publications in the field of EMs. The selected papers were analyzed and reviewed to: (1) list and explain how data-driven methods enhanced energetic material design and discovery, (2) capture the basic concept of the utilized data-driven methods and architectures, and (3) present the remaining challenges of applying data-driven methods to energetic materials and suggest possible directions for future investigations. Noting that the field of AI/ML is rapidly evolving, we acknowledge that this review is not intended to be exhaustive or comprehensive in its coverage. Instead, the review intends to gain an overall perspective on the progress of AI/ML research in the field of energetic materials.

## 1.6 Forward and inverse design

The material design cycle depicted in Figure 1 includes aspects of forward and inverse designs. The forward design seeks to modify the property and performance of the material, by directly manipulating its internal (micro-/meso-) structure. The forward design process can be achieved through experimental evaluation or computational modeling. Experimental evaluation is preferable in some fields of material science where the evaluation does not demand excessive time or resources. However, because experiments on EMs tend to be expensive, laborious, and hazardous, numerical simulation methods are often relied upon for forward design.

Numerical modeling and simulation of EMs can be challenging, as the response of such materials occurs over a short time and small spatial scales, typically within a few nanoseconds in time-scale and over a few micrometers in length-scale, [42-43]. A variety of computational models and well-established codes have been developed and deployed to resolve the response of explosives subject to shock loads [1]. Once they have been rigorously validated, computational forward design models allow us to not only quantify properties and performance but also to understand the underlying mechanics of the ignition and combustion process, such as the crucial hotspot physics and its connection to the microstructure [1, 16]. In recent years several research groups have furthered understanding of the relationships among structure, property, and performance of a range of CHNO EMs [44-47]. These studies have revealed important relationships among porosity [1], pore size [1,48], shape and distribution [15, 45, 49-51], and other morphological features such as asperities and interfaces between crystals and binders [46, 52], and their effects on the sensitivity of the composite material.

As opposed to forward design, the inverse design process aims to find a structural configuration or micro-architecture to meet desired/targeted property and performance QoI. The typical methodology for inverse design involves a domain expert who may suggest, based on a hunch, a candidate microstructure to achieve the targeted property or performance QoI. The advancement of computational models now allows the designer to prototype a microstructure quickly *in silico*, perhaps employing synthetic microstructures that can serve as surrogates for real microstructures. The suggested microstructure is sent to the forward design simulations to be evaluated and the material scientist can determine if the candidate microstructure produces the desired QoI to satisfy some norm of expectation. If the design is not deemed satisfactory, the above steps are iterated until the satisfactory microstructural design is achieved. This "cut-and-try" (experimental/computational) method is not only time-consuming and laborious as it commonly takes hours or days to evaluate a single case, but also requires the close involvement of the domain experts in the design loop. Furthermore, this process may limit the search space due to human bias [53] or legacy knowledge. The vast design space and time intensive process make this route to MbD a challenging problem.

MbD efforts in the broad field of materials science have generally included some common sub-tasks. The perspective on which of those sub-tasks is more important can vary, but three critical subtasks can be identified:
1. Microstructure characterization and representation is a critical subtask as it defines the design search space.
2. Establishing S-P-P linkages is critical to the MbD problem to connect the microstructure to the properties of the material reliably and rapidly.
3. An efficient optimization method is desired for the MbD problem as evaluating all the candidates in the design search space would be impractical.

These three components are common and intrinsic to the MbD process and are depicted in Figure 1. In the following, these three primary aspects of the MbD process are described, along with issues associated with AI/ML approaches and their use in MbD for a wide class of materials, including EM.

## 2. Data representation for microstructure design

Material microstructural representation/design approaches are usually not compatible directly with off-the-shelf machine learning methods. The representation of microstructural data is of critical importance as it defines the search space (or design space) of the materials of interest along with the design parameters to be optimized. Energetic materials, in particular, have complex microstructural morphologies (porosity, shape distribution of voids and crystals) [1] which have a direct impact on the material properties (strength, reactivity) and performance (initiation sensitivity, energy release rate) [1, 45-46, 49, 54]. For example, it has long been known that the number density and size of the voids in an EM affect the sensitivity of the explosives [1, 48-49, 51]. Also, elongated voids and voids oriented in parallel with the direction of the shock tend to be more sensitive [16, 51] than nominally circular voids or voids that are oriented perpendicular to the direction of shock loading. Therefore, in a given microstructure of an EM, the pattern and distribution of voids and other defects in the microstructure play a key role in the shock response of the material. Thus, a method to accurately describe (i.e., how the voids and crystals are distributed in the microstructure, etc.) and control the geometric configuration of the microstructures (i.e., changing the orientation of the voids, etc.) is needed to represent the search space correctly and to effectively and fully execute the inverse design process. The data representation methods discussed in the following subsections are organized in Table 1 with which materials it has been applied to.

| Method | Materials |
|---|---|
| Neural Styles | <ul><li>Class III and class V pressed HMX [44]</li><li>Sandstone [74]</li><li>High carbon steel alloy [44]</li><li>Metal foam structure [44]</li><li>Metal-metal (Ni-Al) composite [44]</li><li>Artificial porous media [75]</li></ul> |
| Autoencoder | <ul><li>Berea [67]</li><li>Sandstone [67]</li><li>Artificial porous media [66]</li></ul> |
| Variational Autoencoder | <ul><li>Plastic-bonded explosive [113]</li><li>Sandstone [73-74]</li><li>Artificial porous media [75]</li></ul> |
| Generative Adversarial Networks | <ul><li>Class V pressed HMX [58]</li><li>Sandstone [57, 61-62, 67]</li><li>Berea [67]</li><li>Artificial porous media [128]</li></ul> |
| Conditional GAN | <ul><li>1,3,5-triamino-2,4,6-trinitrobenzene (TATB) [84]</li><li>Sandstone [73, 83]</li><li>Ultra high carbon steel [80]</li><li>Battery [73,83]</li><li>Silica [73, 83]</li><li>Artificial porous media [81-82]</li></ul> |
| Discrete latent representation | <ul><li>Low carbon steel [76]</li></ul> |

**Table 1. Overview of data representation methods discussed in Section 2 with its application.**

## 2.1 Shape descriptors in idealized microstructures

EM microstructures are commonly obtained as greyscale images through X-ray computed micro-tomography (xCMT) or scanning electron microscopy (SEM) [54]. For many computational approaches, however, the raw pixel-based representation of microstructure is not of practical use as it does not represent the microstructural features (i.e., the distribution of void sizes cannot be easily related by how the pixel intensities are spatially located on the image). Hence, there have been many approaches taken to represent microstructures precisely, i.e., to capture important features (particle size, porosity) observed in real microstructures and to cover the range of feature characteristics to represent various microstructures.

The representation of EM microstructures is challenging as it must be able to capture the complex geometry of voids and crystals. In EM research, it has been a common practice to develop an idealized representation of the microstructure using geometric primitives (i.e., ellipsoids, polygons, etc.), and simple shape descriptors (i.e., the distribution of void/crystal sizes, aspect ratio, orientation, etc.) [15-16, 44-46, 49-50]. Such "idealized" representations of microstructures simplify the search space to keep the problem of MbD tractable. However, these representations may fail to capture complex shapes such as elongated voids or highly branched voids, which are known to have significant impacts on the property of EMs [15]. Bostanabad et al. [23] summarized the conventional methods for microstructure characterization including statistical descriptors (e.g., N-point correlation [55]), and physical descriptors (geometric descriptors such as void size and composition descriptors such as porosity) [54]. To overcome the limitations of idealized representation, Fourier-coefficient-based methods have been applied to represent EM microstructures [50]. The idea is to capture morphological characteristics of microstructures via a frequency transformed representation, in which low frequency components represent the overall shape and high frequency components represent fine details. However, the interpretation of Fourier coefficients is not so straightforward for complex (e.g., branched) shapes, as different geometric features may get entangled within the same frequency. Thus, it is desirable to devise a microstructure representation technique that is not limited to a certain subspace of the microstructural design. Idealized shapes and Fourier-based or other means (e.g., N-point correlations [55]) for capturing the complexity of EM microstructure may limit the design space that can be explored.

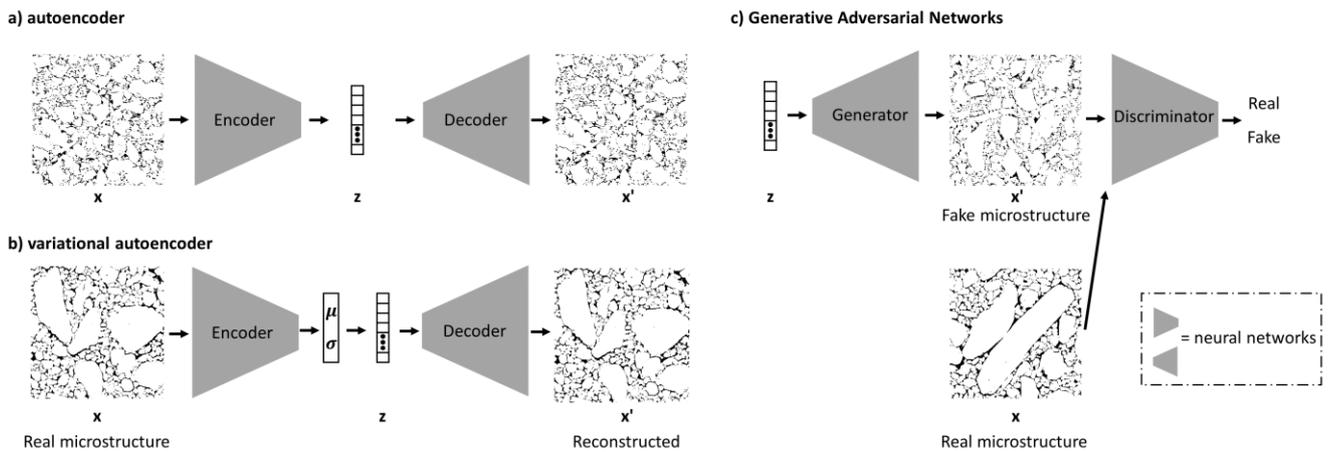

**Figure 2.** The schematic diagram of most commonly used ML generative models. (a) autoencoder takes the microstructure ($x^{h \times w}$) as an input, neural network encoder compresses the dimension to the latent vector ($z^d$) to learn the latent representation of the microstructure, and neural network decoder reconstructs the original image back ($x'^{h \times w}$). (b) variational autoencoder operates similar to autoencoder but learns the mean vector ($\mu$) and standard variation vector ($\sigma$) of the microstructure and use them to sample a latent vector (z) for reconstruction process. (c) the generative adversarial network starts with randomly initialized latent vector (z) and generate a fake microstructure, whereas the discriminator takes real microstructure (x) and fake microstructure (x') as an input and learns to distinguish whether given microstructure is real or fake. It is commonly assumed that the dimension of latent vector (z) is much lower than the dimension of original microstructure ($d \ll h \times w$)

## 2.2 Neural networks for geometry/data representation

We AI/ML approaches can assimilate and provide a richer coverage of shape descriptors. Compared to the traditional approaches above, deep learning methods have some appealing benefits for microstructure representation. The power of deep representation comes from its ability to parametrize complex high-dimensional data (image pixels/voxels) into a low dimensional vector representation with latent variables, called a latent vector [25, 56]. While hand-crafted shape descriptors may be biased toward certain visual traits and may overlook other important features, latent vectors learned via deep representation learning tend to suffer less from such bias. Furthermore, latent vectors can describe complex microstructures with greater precision, compared to traditional shape descriptors in which a significant level of detail is likely overlooked.

In the case of CNNs, imaged microstructures can directly be used in their pixel/voxel form. CNNs can then assimilate microstructural morphology directly from image data through feature maps, or the patterns of neural activations in convolution layers. The idea is that, since these feature maps encode geometric patterns of pixels, they can be used as shape descriptors [44, 57-63]. The benefit of the CNN-based approaches is that the human designer does not need to predefine a set of features to describe shapes. Instead, CNNs can discover useful shape descriptors from data by themselves. This opens the possibility of illuminating previously ignored features as well as reducing human bias. However, these machine-learned features can be difficult to interpret, and their physical meanings may be obscure. The following subsections describe and assess the strength and weaknesses of the different types of deep learning techniques that have been employed for microstructure representation and generation in the context of EM.

## 2.3 Neural styles

Artificial neurons in a neural network produce 'activation values' in response to input image patterns. Different neurons react to different shapes and patterns in the input image. The strength, or magnitude, of activation, indicates how prominent the corresponding pattern is in the input image. Thus, in a neural network trained on EM, there would be, for example, neurons responding to certain elongated voids, or neurons looking for contorted voids. These neurons will activate differently depending on what kind of void shapes are presented to them, and hence, their activation levels may serve as a numerical shape descriptor for EM microstructures.

Gatys et al. [64] built their method upon such an idea, i.e., that the patterns of neural activations may indicate the "style" of the input microstructure image, and therefore, could be used to define a design space for microstructures. In Gatys et al. [64], a CNN was presented with a microstructure image, and its neural activations were recorded in a tensor form. To explore the design space with such neural-encoded shape representation, they introduced the approach called 'neural style transfer,' in which the unique "style" of one image is transferred to another image to generate new images of the same style, but a different variant. In the EM domain, Roy et al. [44] took this approach to extract the style information of Class III and V HMX and generate an ensemble of microstructures that are facsimiles of the given image. The generated microstructure resembled the conditioned style but varied in its content. The technique allowed for the generation of an ensemble of stochastic microstructures, which could serve as specimens for S-P-P linkage simulations [17].

Neural style transfer can perform well at extracting the global pattern of microstructure voids and crystals and can generate multiple synthetic samples of microstructures with a similar visual look. However, new synthetic microstructures can only be generated when there is a reference microstructure to extract the style from. This means that the design exploration must be limited to the proximity of an existing sample, since the neural style transfer algorithms only "mimic" the existing style, but do not interpolate or extrapolate [58]. Also, the boundary of voids and crystals of the generated image tends to be blurry [58]. In addition, the neural style transfer algorithms require training a new model for any new style to be extracted from. These issues can limit in finding of optimal designs for QoI and hinder novel microstructural design.

## 2.4 Autoencoders

Autoencoders [65] are a special type of neural networks. They consist of an encoder network for mapping a microstructure image to lower-dimensional representation (e.g., latent representation) and a decoder network to create a facsimile microstructure image from the lower-dimensional representation. The autoencoder, depicted in Figure 2a, thus has an hourglass-shaped architecture comprised of the juxtaposition of the encoder (from high-dimensional information to lower-dimensional information) and decoder (from low-dimensional information to high-dimensional information). The autoencoders function as unsupervised methods as there is no human supervision or labeled data required. Instead, autoencoders can be trained simply by comparing the original image against the reconstructed image and minimizing their differences. Therefore, compared to other AI/ML methods, autoencoders are less laborious to train, making them a suitable candidate for representation learning in many MbD applications.

An autoencoder is usually trained to produce outputs identical to inputs, which may sound trivial. However, the hourglass-shaped construction of neural network layers creates a "dimensional bottleneck" in the middle of the network, making the problem non-trivial. That is, the middle layer of an autoencoder network has a far smaller number of neurons compared to the input and output layers. For information to pass through such a dimensional bottleneck, the neural network must learn to compress information into a lower-dimensional encoding and to decode such an encoding to reconstruct the original information. Due to these architectural details, autoencoders learn how to effectively summarize data in the form of a low-dimensional vector, often referred to as the latent vector. In the context of EM structure representation, an autoencoder may take a microstructure image represented in a high-dimensional pixel space as input and compress the pixel values into a low-dimensional latent vector. The latent vector is then decoded by the latter half of the autoencoder (decoder), to reconstruct the original microstructure image as output. Upon successful training, the latent vector encodes important geometric characteristics of the input EM microstructure, that is critical to reconstruct the original microstructure image. Hence, the latent vector learned by the autoencoder defines the design space/search space of an MbD problem, upon which S-P-P learning and design optimization problems can be based.

There are several examples of the use of an autoencoder for data representation for material microstructures, including EM. Zhang et al. [66] applied the autoencoder algorithm to learn the latent representation of low-resolution porous media. They showed the efficacy of auto-encoded latent vectors in predicting the permeability of porous materials with $R^2$ of 0.896 and mean squared error of 0.064. This amounted to improved prediction performance by one order of magnitude compared to directly predicting from the microstructure image. Shams et al. [67] applied an autoencoder as part of the multi-scale analysis for sandstone microstructures where autoencoder and GAN learned to generate inter-grain and intra-grain pores of the multi-scale heterogeneous microstructures, respectively. The suggested multi-scale process showed good performance, as evaluated by morphological measure (e.g., range of porosity), statistical measure (e.g., 2-point correlation), and transport property measure (e.g., permeability) and showed good agreement in the range of such measurements between the generated microstructure and samples of the training data set. Nguyen et al. [41] used the U-Net architecture, a special type of autoencoder, as the means for learning a shape descriptor and used it for the prediction of the thermomechanics during the ignition and growth of hotspots in shocked EMs. Solving such a highly complex system would require a deeper architectural design of neural network (with more parameters), but such a deeper architectural design commonly faces degradation of performance [68]. Neural networks are typically trained with error back-propagating from the output layer to the input layer. However, during the back-propagation, the gradients may dwindle to zero or diverge exponentially, depending on the choice of activation functions [69], preventing the neural network from converging. However, unlike other autoencoders, the U-Net architecture contains skip connections, which is essentially a shortcut connection between the encoder and decoder to achieve better flow of the gradients and to propagate spatial information that may otherwise be lost during the dimension reduction process. Li et al. [70] experimentally showed that the loss landscape changes significantly when a skip connection is introduced to the deep neural networks. Thus, the U-Net architecture allows for a deeper neural network, which results in richer hierarchical shape descriptors for the heterogeneous EM where the complex geometry of the voids and crystals (e.g., elongated voids) are learned from the stack of earlier layers and the finer aspects of the pattern of such complex geometry (e.g., the general pattern of size and orientation of voids) of the microstructure is captured from the deeper layers. Nguyen et al. [41] found that the U-Net-learned shape descriptors for EM were effective in predicting the S-P-P linkage and for capturing the thermomechanics of ignition and growth of hotspots in shocked EM.

While autoencoders can provide effective, easy-to-implement means for microstructure representation, they should also be used with caution. First, when autoencoders are not properly regularized, they may lead to representations that are discontinuous and/or incomplete [71]. A continuous representation must guarantee that two points that are close in the latent space yield similar microstructures once decoded. A complete representation must guarantee that the encoded microstructure, i.e., a point in the latent space should produce quantitatively and qualitatively realistic microstructures once decoded. Unfortunately, naïve autoencoder formulations can guarantee neither the continuity nor the completeness of representation. Variational autoencoders (VAE) [72] could be employed to address this problem, in which explicit regularization is achieved via stochastic latent vectors, depicted in Figure 2b. In fact, VAEs have been applied to learn the latent representation for porous media [73-76], which showed significant improvement in their performance. However, in the case of VAEs, or other regularized autoencoders, generated microstructure images tend to be blurry due to the strong regularization of the model [77].

2.1 Generative Adversarial Networks (GAN)
We porous media").

## 2.5 Generative Adversarial Networks (GAN)

Since its first introduction in 2014 [77], GAN has been one of the most influential research topics in the computer vision community, with applications such as human face synthesis, music generation, medical image analysis, and human pose synthesis, to name a few [78-79].

Generative adversarial networks (GAN) refer to neural network algorithms based on the idea of adversarial competition among neural networks. In a typical setup depicted in Figure 2c, GAN is comprised of a generator network that generates synthetic images and a discriminator network that distinguishes those synthetic images from real images. To begin the training process, these two networks are randomly initialized. Through iterations, the generator network attempts to "fool" the discriminator network by learning to generate more realistic synthetic images, whereas the discriminator network tries not to be fooled by developing more accurate discrimination criteria. Chun et al. [58] first adapted GAN for EM research. They introduced a spatial GAN architecture to generate synthetic HMX microstructures, concerning two different types of the spatially defined shape descriptors, namely parameters for the global morphology and parameters for the local stochasticity. The global morphology parameters define the overall "style" of the microstructure, whereas the local stochasticity parameters add a stochastic variation to the microstructure. Based on these shape descriptors, which could also vary spatially across different locations, Chun et al. [58] were able to produce realistic microstructures of pressed HMX. In addition, they also demonstrated fine-grained control of microstructure morphology, in which key morphometry parameters including void/crystal sizes, aspect ratios, and orientations could be meticulously controlled not only globally but also locally. Furthermore, they could also generate diverse, stochastically varying microstructures, while keeping the global morphology the same. Compared to neural style transfer methods, the GAN-learned latent space in Chun et al. [58] was continuous and mathematically compact, such that smooth interpolation and extrapolation of microstructures were possible. Also, the results from the GAN method were crisper and less blurry compared to the results using the autoencoder-based methods.

However, similar to other DL methods, Chun et al.'s work could not overcome the limitation that the latent parameters lacked human-interpretable physical meaning. Although they presented a regression model to map the GAN parameters onto a few known morphometry parameters, their work did not propose a fundamental solution to the problem. To this end, conditional GAN (or conditional VAE) is a variant of GAN (or VAE) that can generate microstructure images conditioned on a physical condition (such as processing conditions [80], species of material [81], etc. [82-84]). The conditional generation is learned through the additional information (e.g., auxiliary information of class "labels" such as different species of material) flowing through the generator and discriminator from the labeled data set, and thus it requires a label for each microstructure image during training. Feng et al. [83] used conditional GAN (cGAN) to generate porous media conditioned on void morphology, while others were conditioned on porosity [81], etc. Meanwhile, Liu et al. [84] tried to extract knowledge of the porous media by including inverse interpretation of the GAN framework, by using attGAN [85]. The attGAN was first introduced in 2019 for human face generation. The key contribution of the attGAN is that it allows editing of the local region of interest while preserving the representation outside of the region of interest. This may allow the designer to generate microstructures with more control of local features such as the void/crystal size, porosity, etc. in a region of interest (RoI).

Recently, the idea of discrete latent representation has become a popular choice to overcome some of the issues of GAN, such as instability in learning [86]. Zhang et al. showed that discrete latent spaces can effectively retain semantically more interesting information and learn more diverse representations due to the regularization and feature matching property [86]. Noguchi et al. [76] applied vector-quantized-VAE (VQ-VAE) [87] to learn the discrete representation (called codebook) of the microstructure image of a low-carbon steel. VQ-VAE adapts the architecture of the VAE and constructs the codebook via vector quantization. The discrete latent representation is an active research area with different quantization methods (e.g., vector quantization [87-88], feature quantization [86], etc.) on different architectures of the ML methods (e.g., VAE [87], GAN [86, 88], etc.). The transformer [89] is one of the state-of-the-art methods in natural language processing tasks (and vision transformer [90] for the vision tasks) where it uses a self-attention mechanism to learn the features from the intra-correlation of so-called "codes", which is a latent representation of microstructure image partitioned with the smaller RoIs. The discrete latent representation (called "codebook") makes the application of the transformer natural as the microstructure is represented with series of "codes" in the codebook. The combination of the transformer and the codebook has shown promise in many fields, including high-resolution natural image synthesis [88], image classification and object detection [91], speech synthesis [92], etc. To the best of our knowledge, the combination of the discrete latent representation and transformer has not been applied to EMs.

## 3. S-P-P linkages

In the design loop, once the material structure is represented and quantified in some way, the next step is to link the structure with the response of the material to loads, i.e., to establish S-P-P linkages. As shown in Figure 1, this is a key aspect of MbD and requires the development of computational and experimental methodologies and tools to quantify the S-P-P linkages. With advances in computational modeling capabilities, it has become increasingly possible to infer properties of materials from microstructures through a large number of in silico experiments. However, due to the complexity of EMs, it often takes hours or days (depending on the size of the simulated sample and the accuracy of the predictive calculations) to evaluate a case even on high-performance computing clusters (HPC) [1, 15-16, 19]. This is where opportunities arise for AI/ML models to speed up the process of S-P-P linkage assimilation. As evidenced by works such as Nguyen et al. [41], AI/ML approaches can aid in reliable and rapid estimation, by suggesting sub-spaces in the overall parameter space where estimation can be improved. They can also indicate promising regions of the design space that may lie outside the human practitioner's natural interest based on past experience or prior knowledge. The various approaches for S-P-P linkages through AI/ML are summarized below and depicted in Table 2.

|   | **Model** | **Structure Representation (Input)** | **QoI (Output)** | **Reference** |
|---|---|---|---|---|
| **Non-DL methods** | Support Vector Machine/Regression | Structural/chemical features | Bulk, shear, and Young's modulus | Balachandran et al. [96] |
|   |   |   | Strain | Yuan et al. [97], |
|   |   |   | Band gap | Lu et al. [102] |
|   |   |   | Thermal hysteresis | Xue et al. [99] |
|   |   | Void size distribution | Run-to-detonation | Walter et al. [110] |
|   |   | 3D microstructure image | Classification of microstructures based on void sizes | Sundararaghavan et al. [112] |
|   | Decision Tree / Random Forest | Structural/chemical features | Material phase | Roy et al. [101] |
|   |   |   | Band gap | Lu et al. [102] |
|   |   | Void size distribution | Run-to-detonation | Walter et al. [110] |
| **Deep Learning methods** | Multilayer Perceptron | Principal components of microcrack size values and impact velocity | Ignition (go/no-go) | Liu et al. [115] |
|   | 3D CNN | 3D microstructure image | Permeability | Hong et al. [119] |
|   |   |   | Classification of microstructures based on void sizes | Sundararaghavan et al. [112] |
|   | VGG | 2D microstructure Image | Volume fraction | Li et al. [116] |
|   | ResNet / wide-ResNet |   | Composite stiffness | Kim et al. [117] |
|   |   |   | Peak stress | Liu et al. [84] |
|   |   |   |   | Palmer et al. [118] |
|   | DenseNet |   |   | Gallagher et al. [113] |
|   | PARC |   | Evolution of temperature and pressure fields | Nguyen et al. [41] |

**Table 2. Overview of different AI/ML methods for S-P-P linkages discussed in Section 3 with the summary of input-output relationshipo of each application.**

### 3.1 Background on ML for S-P-P linkages in material sciences

ML has been applied to model the S-P-P linkages of materials in general for a long time and there exists many successful examples of utilizing ML for the S-P-P linkages. Data-driven methods can be classified broadly based on how they use the data: supervised learning, which requires a pairing of the input data set with a labeled output, and unsupervised learning, which only requires the input data without a labeled output; semi-supervised learning sits in between [25]. The supervised method can be further broken down into classification (e.g., assigning the material to a general type or class of "similar" materials given an input sample [93]) or regression tasks (e.g., prediction of a property, such as the permeability of a porous material [57]). A variety of AI/ML methods that fall under these broad categories have been used for various materials applications.

The Support Vector Machine (SVM) is commonly used ML models with its roots in statistical learning theory since the 1960s [94]. SVM essentially draws a decision boundary to separate the classes of data by first mapping the input data into a higher dimensional space to make the separation easier, with various mappings provided by different kernel functions [95]. The Support Vector Regression (SVR) is a variant form of the SVM for the regression tasks with the same principles hold from the SVM. The choice of the kernel function needs to be made with care, but it has been widely used in the field of material science. For example, Balachandran et al. [96] used the SVM with radial kernel function to predict the elastic properties (e.g., bulk, shear, Young's modulus) of ceramic materials of M2AX-family, while Yuan et al. [97] used radial and linear kernel function to predict the strains of piezoelectric, and many more [98-99].

The decision tree (DT) is another approach that has been applied widely. It is based on a hierarchical decision scheme. The tree-like structure is composed of a root node (with all the data), a set of internal nodes (which splits the data), and a set of terminal nodes (leaves). Thus, a binary decision to separate the data from one class to another is made at each internal node until the terminal nodes or leaves are reached. The random forest is another tree-based machine learning algorithm that contains multiple decision trees [100]. The decision tree and random forest have been applied in many applications of different materials in general, including high entropy alloys [101], hybrid organic-inorganic perovskites [102], and others [103].

Neural networks have increasingly become the popular choice for S-P-P linkage modeling, despite being black-box models [38, 103]. CNNs have been applied to predict shear strength and melting point to name a few but further discussion can be found in the review article by Choudhary et al [43]. The Graphical Neural Network [104] follows a similar architectural basis as the other deep neural networks in general but employs a graph structure as an input and output. Thus, it is a natural fit for applications in material informatics with chemical/molecular compounds where such graph-like connectivity (e.g., molecular bonds) characterizes the dataset. In this context, GNNs have been used to predict the polar surface area, partial charges [105], detonation velocity, detonation pressure, heat of formation, density [106], etc. [37, 105, 107-108]. media").

## 3.2 ML techniques for S-P-P linkages in EMs

ML applied to approximate the S-P-P linkage of energetic materials covers the gamut of methods described above. Many of these methods may only assimilate part of the S-P-P relationship, for example, connecting structure and property or connecting structure and performance. The works that connect all three components, i.e., connecting the microstructure all the way through shock response of the microstructure and then to the macro-scale response are sparse in the literature [41, 109].

Support vector machines (SVM) and random forest (RF) have been applied to the morphometric representation of the microstructure to predict the property of run-to-detonation [110], critical threshold velocity [111], permeability [112], and others [103, 112-113]. On the other hand, autoencoders or variational autoencoders have been applied to learn the latent representation of the raw microstructure images [74-76]. The ML methods for the latent representation can provide better performance than using the morphometric features directly if there are strong correlations among the morphometric features [114]. For example, Liu et al [115] applied principal component analysis to the morphometrical features of the microcracks and further applied simple NN to predict the ignition of a plastic bonded explosive (PBX) material, i.e., connecting structure to performance directly.

ML can be applied to not only learn the direct relationship between inputs and outputs but can be also applied as components of existing numerical simulation models. These approaches feature a lightweight but sufficiently accurate computational model for estimating the S-P-P linkage and will contribute to the acceleration of EM design. For example, efforts have been devoted to developing machine learning techniques to assimilate knowledge and provide closure models for energy localization within EM microstructures. Nassar et al. [48] used a machine-learned model relying on the Modified Bayesian Kriging (MBKG) method in which the reaction progress rate is formulated as a function of loading (shock pressure) and void size. The derived surrogate model is then utilized in Sen et al. [17] in a meso-informed ignition and growth model for multiscale shock-to-detonation transition (SDT) simulation of pressed HMX materials. These works represent the complete S-P-P linkage. The authors also used the Bayesian Kriging approach to capture the effect of non-cylindrical void shapes and void-void interactions on hotspots and ignition growth of shock-initiated EM microstructures. Their model is capable of accounting for the impact of microstructure morphologies, including void aspect ratio, void orientation, and void fraction on the reactive void collapse. The model is not only useful in a meso-informed macroscale simulation model of pressed HMX, but also for studying the dependency of hotspot ignition and growth rate on the morphological metrics. Sen et al. [17] proposed a multiscale framework for the simulation of SDT in pressed energetic materials. Their multiscale simulation model was successful in determining the go/no-go boundaries and identifying the critical energy for SDT with pressed HMX, providing a useful and reliable tool for the characterization of EM sensitivity.

Deep learning based methods are widely applied in the application of the EMs and off-the-shelf deep learning methods, such as multi-layer perceptron, wide ResNet, VGG, DenseNet, and PixelCNN, have been applied to predict the various properties of the energetic materials from the microstructural images, [76, 84, 113, 116-119]. While deep learning methods can provide strong prediction performance, the interpretation of such predictions is challenging as the deep learning methods have the reputation of being "black-box" models [25]. In addition, as noted in Section 3, the scarcity of labeled data sets amplifies the trust issue with such black-box models as the performance of the deep learning models tends to vary on the out-of-distribution of the training data set [103]. To address some of these concerns, Nguyen et al. [41] have developed a novel Physics-Aware Recurrent Convolutional (PARC) Neural Network that can assimilate the mesoscale thermo-mechanics of shock-initiated EM microstructures. Unlike the conventional input-output "black-box" approach of deep learning approaches, PARC is designed to learn the governing equation that describes the thermomechanics of shock-initiated EM microstructure and once trained to make data-driven predictions. Such an architectural design makes PARC physics-aware and interpretable. In addition, following training, PARC is capable of accurate and high-fidelity prediction of temperature and pressure field evolution within shocked EM microstructures. By post-processing, the field evolution predicted by PARC, sensitivity metrics, including hotspot temperature and hotspot growth rate can be derived with an accuracy comparable to that of direct numerical simulation. Thus, the predictions from PARC can be related to physically meaningful information (to the EM expert) such as hotspot ignition and growth rates. Finally, PARC training and prediction can be performed on complex microstructures without simplifying the assumption on the morphology which is not feasible with previous surrogate-based approaches [15-17, 52]. PARC, thus, can serve as an AI/ML "stand-in" for direct numerical simulations and provides the capability of deriving prediction results with multiple orders of magnitude lower computational costs compared to resolved direct numerical simulations (DNS.) With this functionality and speed, PARC is suitable for many tasks in EM materials design that require rapid S-P-P estimation, including inverse microstructure design [24], materials model calibration [1], or in a meso-informed multiscale SDT simulation framework [16, 120].

## 4. Design optimization

The processes inherent to MbD render it not only ill-posed or weakly conditioned but also multimodal, meaning there can be many different types of structures that can satisfy the QoI. The "correct" (physically meaningful, realizable) structure that optimizes the desired QoI needs to be established iteratively, as shown in Figure 1. The selection of an optimization method can vary depending on several factors such as the volume of the initial data set, the size of the search space, the efficiency of the algorithm used for S-P-P linkage, single-objective versus multi-objective optimization, etc. Human-led experimental optimization may constrain the search space due to the introduction of human bias [53]. Thus, an efficient and effective optimization method to navigate the design search space, aided by an AI/ML-assimilated S-P-P linkage is highly desirable to systematically and efficiently reach an optimal microstructural design for the desired QoI. AI/ML algorithms can also aid human designers by providing them with an understanding of the response of a QoI to changes in various design parameters. This can be achieved by "reaching into the black box" and evaluating the information stored in the AI/ML network, e.g., by examining the dependence of the latent vectors in the CNN representation on specific features of the morphology or specific material properties. Preliminary steps in this vein have been taken in recent work on energetic materials [57, 84].

The selection of the optimization method can also vary depending on the design or the goal of the inverse problem at hand. For example, if one desires to find a unique solution for the material design to attain the QoI, a hybrid of genetic algorithms [121] and topology optimization [122] can be applied by first approaching the near-optimal solution and then approaching the optimal solution via a local search. On the other hand, if the sparsity of the training data set is extreme and minimization of expensive DNS calculations is the goal, one can apply Bayesian optimization (BO) [123]. Thus, the choice of the optimization method must consider many factors such as the objective of the design, the sparsity of the training data set, and how expensive acquiring the training datasets from DNS for S-P-P approximation is.

To speed up the S-P-P evaluation, it may be possible in some applications to limit the search space by focusing on a specific objective by using domain expertise, prior knowledge, or understanding of the physics. For example, Lookman et al. [98] limited their search space to only $(Ba_x Ca_y Sr_z)(Ti_u Zr_v Sn_w)O_3$-family of solid solutions and constrained the range of possible configurations with some heuristic rules. In addition, they further limited the meaningful change of the composition to 0.01, which lead to a total of about sixty thousand design candidates. A constrained search space with a computationally inexpensive route to S-P-P approximation allows us to evaluate a larger set of design candidates. However, evaluating all the design candidates in a population pool is not often feasible. Metaheuristic algorithms are a popular choice for global optimization problems. Metaheuristic algorithms are nature-inspired algorithms [124-125] and may provide a sufficiently good solution in a reasonable amount of time by relying on a selected heuristic. Metaheuristic algorithms cover large swathes of algorithms and have been applied to many materials science problems, and the methods used have included genetic algorithms [75, 126], differential evolution, ant colony optimization, particle swarm optimization, etc. [127]. All metaheuristic algorithms use a certain trade-off between randomization and local search, but in general metaheuristic algorithms are suitable for global optimization [125].

Adaptive design methodology, such as Bayesian Optimization [123] or Active Learning [128], is another paradigm for optimization and works especially well in finding extremal QoIs with limited initial data sets. These extremal QoIs can often lie on the out-of-distribution of the training data set [103]. Bayesian Optimization quantifies the balance of prediction from current knowledge (exploitation) and prediction based on the likelihood of observing new knowledge by discovering uncertain regions in design space (exploration) during the optimization and thus can be efficient when choosing the next candidate structure for DNS evaluation. For example, Xue et al. [99] used BO to find the optimal design of NiTi-based shape memory alloy with low hysteresis. They found a set of structures with significantly better performance compared to the initially assumed best performance structure.

Thus, efficient optimization methods are desirable for the inverse design problem, and some common choices of optimization methods and emerging optimization methods applied to the materials in scope are described below and summarized in Table 3.

| Type | Optimization Method | Representation | S-P-P estimation | QoI | Ref |
|---|---|---|---|---|---|
| Gradient-based optimization | Back propagation & gradient descent | GAN | CNN | Stress and Strain (via complience tensor) | Tan et al [130] |
| | | attGAN | Wide ResNet | Peek Stress | Liu et al. [84] |
| Population-based optimization | Genetic algorithm | Microstructure image | ResNet | Composite stiffness | Kim et al. [117] |
| | | VAE | DNS | Maximum temperature & power density | Guo et al. [75] |
| Bayesian Optimization | Expected hypervolume improvement | Conditional GAN | Gaussian Process | Pressure drop and filteration efficiency | Matsuda et al. [81] |
| | Expected Improvement | structural/chemical features | | Adsorption | Deshwal et al. [144] |
| | | GAN | | Porosity, specific surface area, efficiency | Nguyen et al. [57] |
| | Ensemble of expected improvement, lower confidence bound, probability of improvement | GAN | | Energy absorption | Yang et al. [143] |
| ML-based optimization | Actor-critic | GAN | Actor-critic | Porosity, specific surface area, efficiency | Nguyen et al. [57] |
| | Mixture density network | GAN | Mixture density network | Optical absorption | Yang et al. [150] |
| | Invertible neural network (inverse run) | structural/chemical features | Invertible Neural Network (forward run) | Band gap | Zhang et al. [139] |
| | | | | | Fung et al. [140] |

Table 3. Overview of the design optimization methods discussed in Section 4. Note that acquisition function of Bayesian optimization is specified for different applications discussed.

## 4.1 Gradient-based optimization

To overcome the challenge of vast design space, some constraints can be placed on the search space to practically evaluate all designs in a reasonable amount of time. For example, Walter et al. [110] set a specific range of the void size distribution that covers the realistic range of the microstructures. They identified the optimal design of microstructure in terms of a void size distribution that provides a target performance of shock-to-detonation for high explosives.

Although deep learning methods have the reputation of being black-box models, they are differentiable through back-propagation [129]. Thus, some works have applied a gradient-based topology optimization directly to the trained deep learning models. For example, Tan et al. [130] first learned the deep latent representation of porous media via GAN and then modulated the discriminator to predict the QoI. They verified the performance by finding microstructures for various QoIs through back-propagation and were able to find microstructures that resulted in targeted QoIs within 5% relative error. Liu et al. [84] took a similar approach to solve the MbD problem for TATB explosives with the QoI of peak stress. They applied attGAN [85] to learn the latent representation of the TATB microstructures and parametrize with more meaningful parameters such as void size distribution, porosity, etc. to allow for editing local regions of the generated microstructure. In addition, wide ResNet [131] was applied to predict performance QoIs given a microstructure image. They framed the inverse optimization as finding minimal changes in the morphometric attributes that result in the targeted QoI. A relaxed formulation of the objective function with a differentiable Generator and differential S-P-P approximator allowed for the use of back-propagation for calculating gradients. Then, a simple gradient descent method was used to find the configuration of the microstructure and the corresponding attributes to meet the targeted QoI [84].

The gradient-based optimization using neural networks can be appealing as it can find the optimum solution with fewer design evaluations. However, this method requires the robust and reliable performance of the trained neural network model (treated as a surrogate for the "real" system) as the solution is found directly by differentiating the trained model, which may not work when there is a limited amount of data (e.g., Tan et al. [130] used 38k and Liu et al. [84] used 53k training samples). In addition, this approach may not be suitable for finding extremal QoIs as it may not consider many local optimal points. In fact, Guo et al. [75] showed that gradient-based optimization methods have a tendency of finding local optima.

## 4.2 Population-based optimization

Gradient-based optimization methods confine the search space to a certain basin of convergence, which may not be desirable for the purpose of a broader exploration of designs. Furthermore, gradient-based methods are sensitive to local optima. Sometimes, instead, an efficient global search method may be more desirable for MbD of EM.

Population-based optimization methods [124] are a family of stochastic optimization methods that do not rely on gradients for the search for optimum design (gradient-free optimization). Many methods in this family are inspired by how nature goes through the optimization process. For example, ant colony optimization [132] is inspired by the way ants communicate in nature, in which pheromones are used to direct each other to resources while exploring their own environments. Similarly, artificial bee colony optimization [133] mimics how honeybee swarms locate food sources through collaboration. In the sense that these algorithms employ a "population" or "swarm of agents, they are often called population-based search or swarm intelligence methods.

In a similar vein to population-based search, evolutionary algorithms form a subfield of gradient-free algorithms inspired by the Darwinian evolution in biology, conceptualized as "survival of the fittest" [134]. Evolutionary algorithms, such as genetic algorithm [121] and differential evolution [135] start with a set of "chromosomes" in the population representing a candidate solution to the problem. Under an objective function, the population is then evaluated on how "fit" they are. The least fit candidates are eliminated from the population and replaced by new "offspring" chromosomes, which inherit or mutate from the previous generation. The process repeats for multiple generations until convergence.

In materials research, Kim et al. [117] applied a genetic algorithm to maximize composite stiffness. They navigated the vast design space of 160,000 unique microstructures by employing chromosomes evolving through crossovers and mutations, which finally led to an optimum composite with maximal stiffness. The algorithm was able to accomplish such a task within 4 hours of computation time, which involved the evaluation of about 400 microstructure designs suggested by the genetic algorithms. The microstructure design obtained through the genetic algorithm exhibited a stiffness of 1144.7 MPa, which is near-optimal compared to the theoretical upper bound of 1223.7 MPa. Such use cases of gradient-free algorithms can be found in many other applications, which has been reviewed and summarized by Liao et al. [136].

Despite their popular use in MbD applications, population-based optimization methods may be limited in some cases when the design space has a large number of design parameters [57]. Many design parameters would require a larger population size for convergence [137-138], but a large population comes at a higher computational cost to obtain more evaluations. In addition, gradient-free optimization does not guarantee an optimal solution and may produce only near-optimal solutions [124] due to the stochastic process to efficiently navigate the search space. For this reason, hybrids of the population- and gradient-based optimization methods are often employed. For example, Guo et al. [75] applied a hybrid of genetic algorithm and gradient descent method for a multi-objective microstructure design problem of a heat conduction system. Using a latent representation obtained using VAE, the algorithm discovers optimum designs that minimize the maximum temperature and maximize the power density. The hybrid method was not able to find all known non-dominated (optimal) points but was able to identify many new non-dominated designs beyond attainable designs estimated by a training data set. Different hybrid algorithms (e.g., particle swarm optimization and genetic algorithms, conditional GAN and gradient descent, etc.) have been applied to materials in general and showed good performance in finding microstructural designs for targeted QoIs [75, 139-140] and extremal QoIs [115]. These hybrid methods require more evaluations compared to the gradient-based methods but can find near global optimal design more effectively than gradient-based methods [75].

## 4.3 Bayesian optimization

One of the challenges in population-based methods is that the objective function must be evaluated for many populations, which can be costly, particularly in MbD applications. Bayesian optimization (or adaptive design) methods aim to minimize these expensive evaluations by keeping track of the probability of discovering an optimum solution. A typical Bayesian optimization algorithm is comprised of a Bayesian model (i.e., Gaussian processes [141]) to build a surrogate as a Bayesian approximation to the forward process (i.e., S-P-P linkage) and an acquisition function to quantify the likelihood of observing a new optimum design. The acquisition function can control the trade-off between exploitation (i.e., searching for a better optimum near the currently known optimum) and exploration (i.e., broader search in unexplored regions). In-between exploitation and exploration, the acquisition function predicts the next set of design candidates that are most likely better than the currently known optimum. Thus, Bayesian optimization reduces the number of objective function evaluations and is most suitable when there is a limited amount of data or when the evaluation of an objective function is extremely expensive (e.g., through highly resolved DNS).

For example, Nguyen et al. [57] used Bayesian optimization as a benchmark for the design of 3D microstructures of porous media, in which the objective was to achieve target QoIs including porosity, specific surface area, and efficient permeability. They used a latent representation obtained using a 3D GAN method and constructed a Gaussian process model as a surrogate. The expected improvement (EI) function [142] was used as an acquisition function, through which they could converge to optimum microstructure designs in a small number of iterations. Similarly, Yang et al. [143] optimized 2D porous media for energy absorption using a Bayesian optimization method with a latent representation learned by 2D GAN. In their case, however, an ensemble of different acquisition functions, including EI, probability of improvement, and lower confidence bound, were employed. In fact, Bayesian optimization has gained in popularity recently in many applications of materials design, including the design of porous materials [81,144], digital rocks [82], drug discovery [145], and others [96-99].

## 4.4 Neural networks for optimization

While the Bayesian optimization is desirable when evaluation of the data is expensive but is often limited when scaling to the high-dimensional features [146]. There are specific types of the neural networks more suitable for optimizing the microstructure in the realm of the MbD to overcome such challenge.

Reinforcement learning (RL) is a specific type of ML, in which an ML algorithm is trained based on feedback on its action. During the training, an RL algorithm makes an action for a given set of observations. The algorithm is rewarded when the action eventually leads to a better future outcome or is penalized (negatively rewarded) when the action leads to a worse future outcome. Such feedback on the expected future reward is then used to update the ML parameters of the algorithm, and the process is repeated until the algorithm develops an optimal policy and/or value criteria for its actions. Once such a policy or value criteria is developed, an RL algorithm can suggest actions for each given situation that can maximize the expected future rewards.

RL algorithms were originally used for building AI agents, such as an algorithm playing the game of 'Go' against human players ('AlphaGo' [147]) or autonomous (air/water/ground) vehicle systems [148]. However, recently, there has been efforts to use RL algorithms as an optimization solver [149], in which RL is seen as a learning-based heuristic search. The main hypothesis is that once trained on a set of optimization problems, RL can learn a policy to efficiently generate solutions for similar but unseen problems [57]. In fact, traditional optimization approaches need to apply to a general class of optimization problems, such that no problem-specific patterns and trends need to be taken into account. However, for a specific set of problem instances (e.g., designing material microstructures), there may exist similar patterns and trends in the solution space, such that solvers can exploit these patterns from historical data. In this sense, RL can be an effective optimization solver, as it can learn to develop a policy to maximize the expected future reward from a set of similar optimization problems.

In the context of MbD, an "action" of an RL algorithm can be considered as a design modification, whereas a "reward" is a function indicative of how close the designed QoIs have become to the target QoIs as a result of such an action. During training, an RL algorithm may make many design changes (actions) and observe how the QoIs change by trial-and-error. From such observations, the algorithm discovers trends of how the output QoI changes in response to the change of input design parameters. In deployment, given targeted QoIs defined by the design problem, RL should be able to make efficient changes of the design parameters and quickly converge to the targeted QoIs.

Recently, Nguyen et al. [57] applied an actor-critic RL model to perform an inverse design of 3D porous media for targeted porosity, specific surface area, and efficient permeability. In their actor-critic formulation, the "actor" was a neural network that predicts the most effective design change for a given (current) design and target QoIs. The "critic," on the other hand, was another neural network that evaluated how effective such a design change was, based on which the neural network parameters of both the actor and the critic were updated. In their formulation, 3D GAN was employed to represent and synthesize 3D microstructures, and the actor directly interfaced with the 3D GAN to generate synthetic microstructure designs. During training, QoIs were then evaluated on the microstructure designs, and their difference with the target QoIs informed the critic network. Once trained, the actor network, alongside the 3D GAN algorithm, was deployed to design microstructures for user defined QoI requirements.

As Nguyen et al. [57] point out, the RL-based design optimization approach can be more desirable for MbD applications, as the algorithm can learn from similar optimization problems. This can be particularly useful for scenarios involving repetitive design tasks in which the material designer needs to find new designs for each instance with different target QoI. The learning-based approach could also be useful when historical data on similar but different material species are available. In fact, Nguyen et al. [57] show that RL could converge to an optimum design with a significantly lesser number of evaluations than what traditional solvers (e.g., Bayesian optimization) need.

Despite such advantages, however, RL requires a large number of training episodes or a "gym" environment in which the algorithm can efficiently perform trial-and-error experiments. Furthermore, the training of RL is not necessarily straightforward, as RLs are widely known to be significantly more difficult to train than other machine learning algorithms [57].

In a similar spirit to the RL-based approaches, there is ongoing research on the use of neural network predictions for solving design optimization problems. For example, Yang et al. [150] learned a latent representation of porous media using patch-GAN and used a mixture density network (MDN) to learn a direct regression from a QoI (optical absorption) to the latent vector. MDN is a neural network to predict probability distributions of outputs conditioned on an input. Yang et al. showed that MDN could directly map a target QoI to the corresponding design representation with less than 1% of residual error percentage (REP).

Similarly, Zhang et al. [139] employed an invertible neural network (INN) to solve an MbD problem of complex metal oxides ($SrTiO_3$) as a direct regression task. Unlike other feedforward neural networks, INNs learn a bijective relationship between inputs and outputs, such that both forward and inverse mapping could be computed. Candidates of metal oxides for the targeted QoI (e.g., band gap) are produced instantly using learned inverse mapping via INN. Zhang et al. [139] further applied a gradient descent method to enhance suggestions. In addition, Fung et al. [140] proposed an inverse design framework (MatDesINNe) that uses an INN to find a near-optimal solution, filters out noisy candidates (e.g., outside of design parameters in concern), and then performs a gradient descent search to find the optimum. They validated their framework on the task of designing 2D semiconductor materials, e.g., $MoS_2$, for targeted band gap values.

Compared to other iterative search methods, these direct regression methods can suggest near-optimal solutions rapidly (within a few seconds), which can then be utilized as a seed for a more accurate search as in Fung et al. [140]. However, these regression methods require a large number of training data. For example, five thousand labeled data was used to train the MDN by Yang et al. [150]. This raises a practical concern for EM applications, in which such a large number of labeled data is rarely available. Furthermore, compared to gradient-free optimization methods (e.g., population-based search, Bayesian optimization, reinforcement learning), these direct regression methods cannot explore the design space effectively, since they rely on a direct, pre-determined mapping between QoIs and design parameters.

## 5. Challenges and potential directions

In the above sections we discussed various types of approaches for MbD, with their pros and cons analyzed in the context of three critical subtasks, viz., data representation, learning S-P-P linkages, and design optimization. In the following, we highlight some of the remaining challenges of applying AI/ML methods to EMs and suggest potential directions for research.

### 5.1 Data scarcity

Thermo-mechanical responses of EM microstructures involve complex interactions between the collapse of individual voids, phase transitions, crystal-crystal interactions, and crystal-binder and crystal-inclusion interactions at interfaces in the microstructure. Short time scale molecular processes also contribute to the evolution of the temperature and pressure, which develop over a few nanoseconds and on spatial scales of micrometers. State-of-the-art direct numerical simulations can accurately resolve these phenomena, but calculations under the continuum approximation can be computationally intensive. Thus, the acquisition of labeled datasets for energetic material microstructures evolving from insult to initiation and beyond can be expensive. The consequent limited datasets add an additional layer of challenge for data-driven methodologies, as the rates and accuracy of learning for most "big data" reliant ML models are driven by the quality and quantity of the available data. Also, the inherent stochasticity of microstructures requires an ensemble of evaluations, but intensive computations (not to mention experiments) over sufficiently representative sample sizes of microstructure are a process-bottleneck even if one can come up with a good design strategy. In the subsections below, we introduce potential research directions and challenges to overcome such issues of data scarcity.

*5.1.1 Transfer Learning: Scaling AI/ML models to different operating conditions*

In the context of EM performance, molecular structure, microstructure, and loading conditions are all coupled together to determine sensitivity and performance. While an MbD paradigm can be designed for one material, over a range of microstructures, it is desirable to not repeat the entire training process for a different set of operating conditions (e.g., shock loading). The sensitivity of the heterogeneous explosive is measured through run-to-detonation distances or James envelopes with the imposed shock pressure information playing a key role as an experimental control parameter [1, 49]. However, it is seldom the case that the modeler has access to the sensitivity data obtained from experiments as well as the corresponding detailed microstructural information. Therefore, establishing S-P-P linkages must rely primarily on high resolution numerical simulations on microstructures, i.e., meso-scale simulations [17]. However, resolved simulations on sufficiently large ensembles of stochastic microstructures for a range of shock pressures is a largely time-consuming process. Thus, supplying AI/ML input datasets spanning the possible loading regimes and microstructures is highly challenging. In general, the modeler also does not have the luxury of preparing labeled data for multiple pressure values. Therefore, the AI/ML algorithm must be able to learn from sparse datasets with regard to the shock loading parameter. Based on this assimilated information predictions must be made for the wider range of shock conditions. This requires an efficient and accurate way to predict behavior in a wide parameter space with sparse coverage of the space by the training dataset. To overcome this difficulty, an adaptive or multi-fidelity [151-155] ML approach may be required. One of the common methods to speed up predictions at unknown points in the parameter space, using knowledge obtained at sparse locations, is transfer learning [156-157]. For example, a neural network trained on one condition of the pressure loading (e.g., at one pressure [41]) could be transferred and fine-tuned on different pressure regimes with a handful amount of data. This approach demands data from simulations only where necessary and adaptively guides the time-consuming input data generation from simulations. The combination of such transfer learning and multi-fidelity approaches can aid in overcoming the anticipated ongoing problem of sparse data availability for energetic materials.

*5.1.2 Meta-Learning: Transfer from one EM species to another*

Transfer learning and multi-fidelity-based methods may be necessary to tackle the scarcity of data that will likely continue to plague EM modeling. In addition to shock loading, there can be another need for transferability, i.e., between material species. Heterogeneous explosives formulated with various molecular species may exhibit similar hotspot physics but differ in how sensitive they are at a given shock pressure, e.g., insensitive (e.g., TATB) to highly sensitive (e.g., PETN) materials [11-13]. Even obtaining a handful of labeled mesoscale training data through direct numerical simulations for one type of explosives for one pressure value takes a considerable amount of time and effort. Thus, a mechanism to transfer knowledge from the trained AI/ML model for one species to another is attractive as it offers a way to "generalize" knowledge across a set of related species of energetic materials, for example, the set of commonly employed CHNO materials. A meta-learning approach [158] can be applied to EMs to extract the "common" knowledge among different species of EMs and the model can be enriched with a limited amount of data generated for the new, as yet unrepresented species. This approach represents a multi-fidelity approach along the species axis in parameter space. In other words, the modeler can use available knowledge on one (or a few) species, transfer the knowledge to an as-yet unfamiliar or "new" species, and refine or enrich the model through a small set of simulations conducted on the new species. This approach may allow for significant reductions in training time to prepare an updated S-P-P approximator for the new species as the network already has some knowledge about the underlying system. In a similar vein, Lansford et al. [108] applied the concept of co-training to the energetic materials in molecular level which utilize similar but larger data set in size to reduce the performance limitation from the scarce EM data.

*5.1.3 Semi- or weakly-supervised learning: speeding up by learning from comparatively cheaper data*

EM microstructures are comprised of voids and crystals of diverse shapes and sizes. High-fidelity mesoscale simulations on these microstructures are highly time-consuming even in 2D, and 3D simulations are unlikely to become run-of-the-mill in the immediate future [17, 159]. An approach to employ the dynamics of sub-elements of the microstructure into the training process can speed up the assimilation by the AI/ML process. The dynamics of single voids, interfaces, and other defects can be computed separately; these sub-elements of the overall microstructure are much less computationally inexpensive. Thus, in this vein, it would be useful to investigate how to learn the collapses of individual voids and their resulting hotspot formations and to connect such local phenomena with global phenomena in a full microstructure.

The evolution of temperature and pressure of explosives under the shock loading can then be treated as an aggregated effect of the contributions of single voids and other features. Nguyen et al. [41] showed that specific configurations of the microstructure are more sensitive to the shock load through single void simulations. Thus, mesoscale hotspot physics can be simulated by aggregating from a collection of the single void physics [16, 45, 48, 159-160]. This approach is tantamount to another route to multi-fidelity modeling of the overall response of microstructures to shock loading.

## 5.2 Uncertainty quantification (sources and methods)

We Machine learning and in particular deep learning are data-driven approaches and demand an adequate set of diverse data points to learn features from the data. However, data scarcity is a common and well-known problem in materials science in general but especially with energetic materials. The performance of the trained deep learning models is less robust on out-of-distribution data points [103], and this generalization issue can be more likely with the limited data set. To avoid the incorrect design of the microstructures of high explosives due to inaccurate predictions from the S-P-P approximator, uncertainty quantification is imperative.

Uncertainty quantification is not only useful for robust prediction but also helps in the adaptive design and optimization as noted in Section 4.3, which can guide exploration in subregions of interest and avoid unnecessary and expensive DNS evaluations. In addition, uncertainty quantification allows to build more trusted data-driven system as the ML/DL models can indicate their certainty on the prediction. Thus, an additional layer of interpretation of the "black-box" model can be achieved [161-162]. In the case of energetic materials, the uncertainties are both aleatoric (e.g., stochastic microstructures), and epistemic (uncertainties in material models' forms, properties, etc.). The predictive model should provide quantification of uncertainty in the parameter space due to both microstructural stochasticity as well as other sources, including the uncertainty in the training data due to numerical approximations in solving the governing equations. These various sources of uncertainty are challenging to capture and quantify due to the sparsity of datasets in practical AI/ML models of EM. In literature, the most common methods appear to be Monte Carlo (MC) Dropout [163], bootstrap sampling (or deep ensemble) [164], and Bayesian Neural Networks [165-166]. Each of the different methods has its advantages and challenges, such as MC Dropout being the most non-invasive, deep ensemble being conceptually the most simplistic and most reliable when tested on the benchmark [167], but requires more time to train multiple models, etc. [168]. In this regard, Abdar et al. [168] provide a thorough review of different uncertainty quantification methods.

## 5.3 Interpretability of data-driven shape descriptors

With the advances in deep neural networks the power of these methods to assimilate complex information regarding material microstructures and their behavior have become an active area of research. However, the new toolkit of DNN carries the reputation of being a "black box", offering little understanding of physics and lacking interpretability. Designers can benefit from "reaching into the box" to obtain information, not to mention insight, into the behavior of the system being designed. The latent search space of the microstructure learned by deep representation learning (such as VAE or GAN) is one of the most popular methods for search space representation. However, understanding of the latent space is not trivial. Conditional Generative Adversarial Networks [169], style-vector [170], and their variants are some of the most common methods to interpret the latent space with controllable generation. For example, Liu et al. [84] applied attGAN [85] for porous materials with the morphometry control variable, such as porosity, void size, etc., where they could smoothly change the specific local area of the generated microstructure while keeping the rest of the images minimally changed [84]. The interpretation of the latent space in terms of the morphological features is highly challenging, but there are ongoing efforts to connect the dots by mapping the latent space directly to the metric space [171], and other physically meaningful representations [172].

In addition, one of the most common limitations of a GAN method is the mode collapse [77, 173]. Mode collapse is when the generator of GAN learns only a small set of output types (such as learning to generate microstructures with only big crystals, but not with small crystals). The Wasserstein loss [173] and unrolling [174] are commonly used to avoid mode collapse. Measuring the model collapse of the GAN models may be straightforward if the data set has a label or perceptual differences in human eyes. However, this is usually not the case for EMs. For example, Chun et al. [58] applied patch-based GAN to learn the underlying distribution of the class V cyclotetramethylene-tetranitramine (HMX) pressed EM. They showed that learned representation covered well the real microstructure distribution of the HMX by indirectly showing the correlation between various morphometry measures (void size, aspect ratio, and orientation) and GAN-learned parameters. However, such correlations show only a few projected relationships, not necessarily the comprehensive explanation of the GAN-learned parameter space.

## 5.4 Continuous convolution due to highly dynamic system

The nature of the simulation data that serve as inputs to the AI/ML algorithm (i.e., the evolution of temperature and pressure field under shock load) is highly dynamic and transient, with the possibility of propagating discontinuities (interfaces) and large gradients (shocks, reaction fronts) [1, 159]. For example, hotspots in microstructures are key to the performance of heterogeneous explosives. Intrinsically, there is a large temperature difference between the hotspot area (i.e., ~4000K) and the non-hotspot area (i.e., ~300K). Therefore, the evolution of hotspots leads to dynamics that displays discontinuities in time (ignition) and space (sharp burn front propagation). Given the dynamic and non-smooth fields to be learned, discrete convolution may not well suited to learn the features of these fast-changing intensity fields with high contrast values (edges, shocks). Nguyen et al. [41] noted that in learning the evolution of temperature and pressure fields using convolutional neural networks, while the location and overall motion of hotspots are relatively well captured, the boundary of the evolving hotspot is not as crisply captured. In computer graphics and animation, similar issues exist of accurately evolving sharp features. Karras et al. [170] proposed the idea of continuous operation of convolution for human face generation on such contrastive regions. They showed a remarkable improvement in generating fine details of facial features and the ability to represent fine details of thin structures like human hair. Adaptation of such approaches to learning from field data generated by flow field simulations through continuous convolution may be able to capture the dynamics of shocks and reaction fronts with higher fidelity.

### 5.5 "Process"-Structure-Property-Performance linkage

In this review, we only considered the S-P-P linkage but excluded the Process in the Process-S-P-P linkage. However, to fully close the loop of MbD, i.e., to actually manufacture the design proposed by the AI/ML design assistant (Figure 1), we must consider the manufacturing process as well. Various sophisticated MbD methods could be applied and suggest a microstructural image for the solution, but there is a potential language barrier between the suggested image (delivered as a rasterized image for example) and how the manufacturer or formulator of the material would be able to process and synthesize such microstructures. Thus, there needs to be another layer of communication in the design loop, where the AI/ML algorithm suggests a microstructural design through data-driven material-by-design but in a language/format that can be utilized by different processing methods. This gap between in silico design and the manufacturing process or shop floor needs to be bridged in future work.

## 6. Conclusions

This review summarized the major components of the MbD problem, namely search space representation, structure-property-performance (S-P-P) linkage, and optimization of the microstructure to achieve targeted design QoIs. Applications of AI/ML to various MbD problems were discussed, with a focus on EM-specific requirements and challenges.

AI/ML for the MbD of EM microstructures is still in its infancy but shows great potential for performance, efficiency, and practicality. As highlighted in this review, a single, universal method that works best for all the problems does not and likely may not exist to cover the overall MbD capability for EM [103]. We should carefully equip ourselves with the AI/ML methods that can tackle the complexity of the microstructure and its dynamics, the design of the inverse optimization problem, the amount of available data (typically sparse), etc. The issue of the sparsity of data, and therefore large uncertainties in the datasets is unlikely to be resolved using current approaches only in the immediate future, particularly where EMs are concerned. Therefore, a suitable AI/ML approach to represent microstructures, assimilate S-P-P linkages, or to optimize the design must contend with sparse data and large uncertainties. Furthermore, given the rather large number of parameters and wide parameter ranges involved, techniques such as transfer learning and multi-fidelity approaches need to be considered to speed up the data-driven design process. Finally, there is a gap between the AI/ML algorithms and computational capabilities and the actual formulation and manufacturing of EM, which needs to be closed before the data driven MbD loop can be realized.

It is anticipated that the combination of improvements in material models, computational methodology, software and hardware, and rapidly evolving AI/ML methodologies is likely to make in silico design of materials, including EM, a strong driving force for developing novel, precise, and tailored materials in the coming decades. Whether this potential is realized will hinge on collaboration between process engineers, advanced manufacturing practitioners (especially 3D printing/additive manufacturing), experimentalists and data-driven modelers.

*Acknowledgments*

This material is based on work supported by the National Science Foundation under Grant No. 2203580 and by the U.S. Air Force Office of Scientific Research (AFOSR) Multidisciplinary University Research Initiative (MURI) program (Grant No. FA9550-19-1-0318; PM: Dr. Martin Schmidt, Dynamic Materials program).